\documentclass{PoS}

\usepackage{subfigure}

\title{MEM imaging of multi-wavelength VLBA polarisation observations of Active Galactic Nuclei}

\ShortTitle{MEM imaging of multi-wavelength VLBA polarisation observations of AGN}

\author{\speaker{Colm Coughlan}\\
        Department of Physics, University College Cork, Ireland\\
        E-mail: \email{colm.coughlan@umail.ucc.ie}}

\author{Denise Gabuzda\\
        Department of Physics, University College Cork, Ireland\\
        E-mail: \email{d.gabuzda@ucc.ie}}

\abstract{We have developed a C++ implementation of the Maximum Entropy Method (MEM) suitable for deconvolving VLBI polarisation data. The first results of this implementation are presented and compared with CLEAN-based deconvolutions of the same data. We present Faraday rotation measure and intrinsic polarisation maps of AGN which have been made from MEM deconvolutions of multi-wavelength observations of Stokes parameters I, Q and U. The advantages of using MEM are demonstrated, in particular its enhanced resolution over the CLEAN algorithm.}

\FullConference{11th European VLBI Network Symposium \& Users Meeting,\\
		October 9-12, 2012\\
		Bordeaux, France}

\begin{document}

\section{VLBI observations of astrophysical objects}

\begin{figure}[h!t]
  \subfigure[Projection onto UV plane]{
    \includegraphics[scale=0.3]{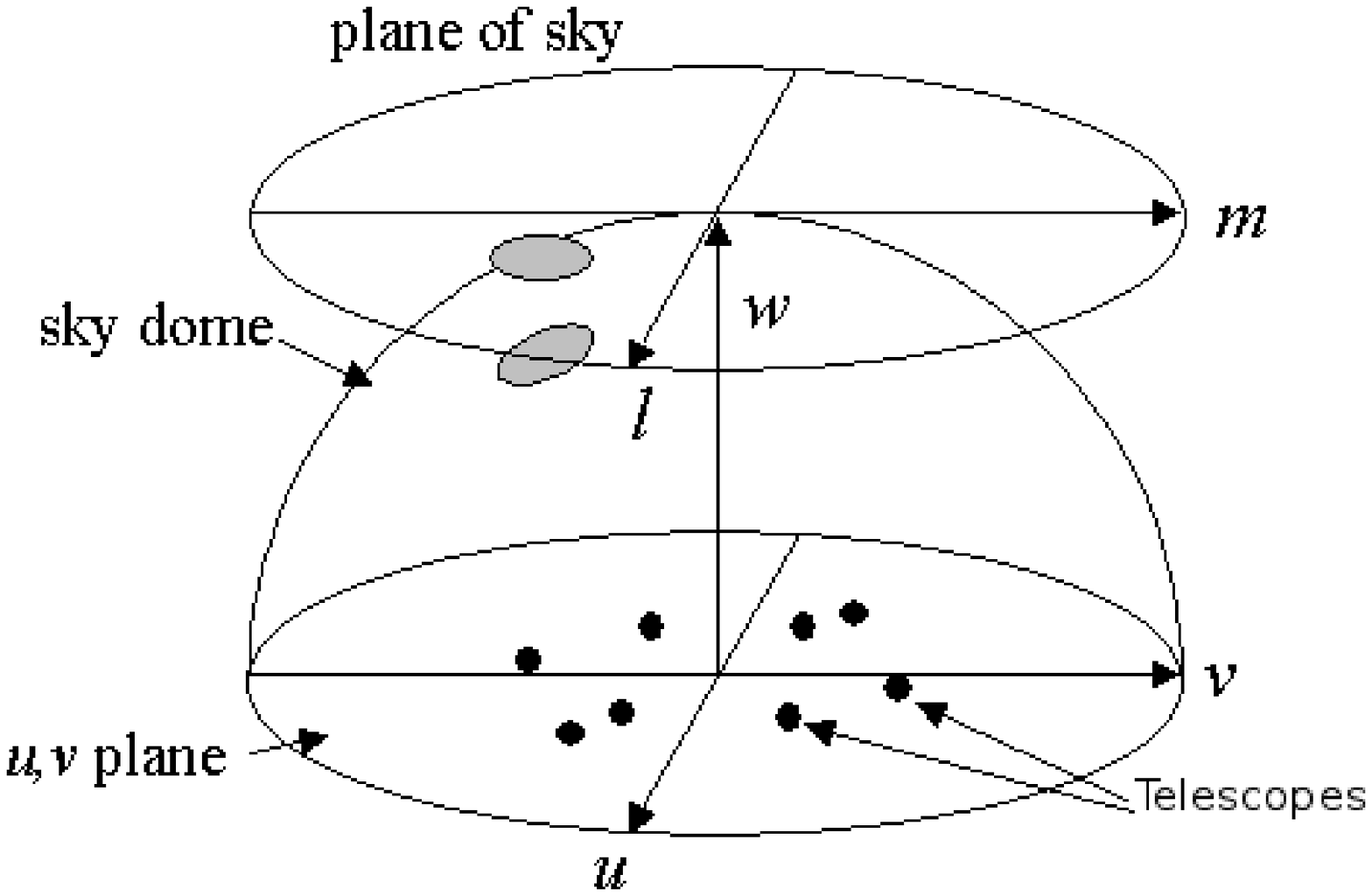}
    \label{uvw_plane}
  }
  \hfill
  \subfigure[UV sampling function]{
    \includegraphics[scale=0.3]{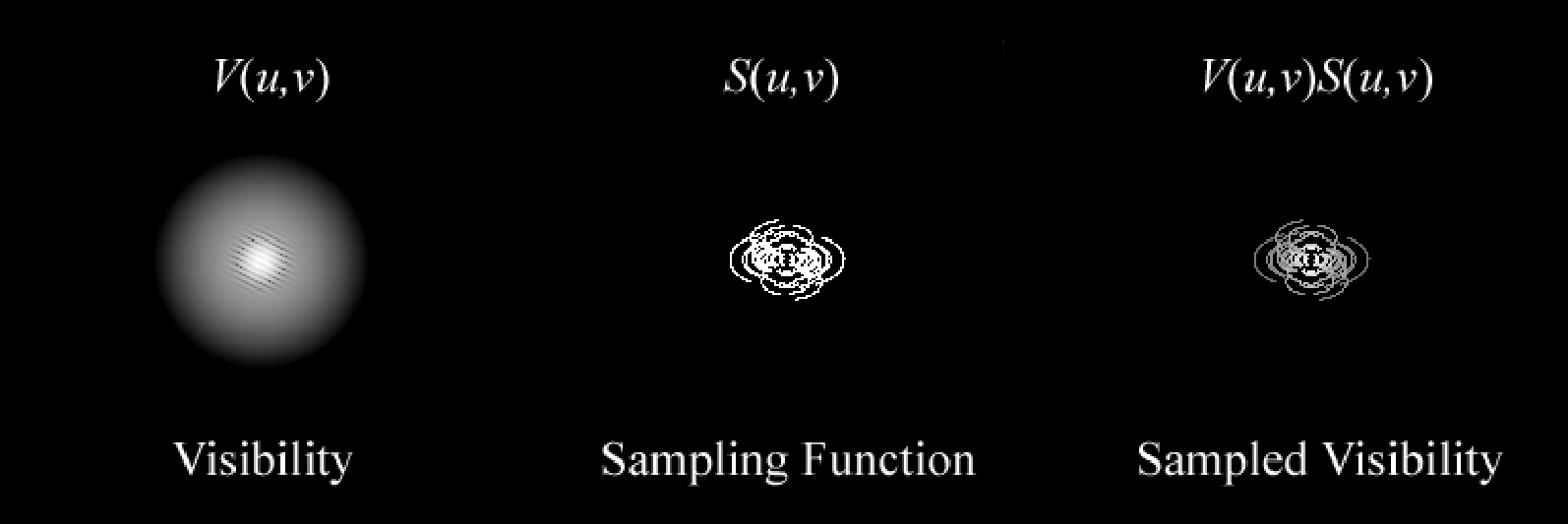}
    \label{sampling}
  }
  \hfill
  
\caption{There is a Fourier Transform relationship between the brightness distribution in the sky and the observations made by ground based telescopes. Only a small fraction of the visibility function is actually sampled. Image : \emph{http://web.njit.edu/~gary/728/Lecture6.html}.}

\end{figure}

Astrophysical objects such as Active Galactic Nuclei have some brightness distribution in the plane of the sky. The output of radio interferometers such as the Very Long Baseline Array (VLBA) actually correspond to the Fourier transform of the brightness distribution, called the visibility function $V(u,v)$ where $u$ and $v$ are the projected baselines (Fig \ref{uvw_plane}).\\

\noindent
Each observation time corresponds to a specific set of baselines $(u,v)$ at which the visibility function is sampled. This situation is improved somewhat by the rotation of the earth, but at the end of an observation one is left with an small sample of the Fourier transform of the brightness distribution one actually wishes to observe (Fig \ref{sampling}). To image the intensity distribution one must take an inverse Fourier transform of the sampled data, however due to the large number of unsampled visibilities, the resulting image will be incomplete, or ``dirty''.\\

\noindent
In order to try to reconstruct the missing visibilities, the dirty map is deconvolved. The product of the sampling function $S(u,v)$ with the visibilities $V(u,v)$ in Fourier space corresponds to a convolution of the ``true'' intensity distribution with the dirty beam in image space. The CLEAN algorithm is one of the most common and effective deconvolution algorithms used in radio astronomy, however other algorithms, such as the Maximum Entropy Method (MEM) are also widely used. It can be advantageous to use multiple algorithms when deconvolving a dirty map, as issues with one algorithm may not be present in another algorithm.

\vspace{-5mm}
\section{The Maximum Entropy Method}

The MEM is an example of a deconvolution algorithm which uses a regularisation parameter to stop the model of the source from converging to the dirty map. If one begins with a flat model source, with total flux equal to the zero-spacing flux observed, and iteratively changes the model until its visibilities agree well with the observed visibilities, the model will begin to resemble the real brightness distribution. However as one does not have a full set of visibilities, if one lets the convergence of the model go too far, it will simply converge to the dirty map. Thus an additional parameter is needed to stop it from converging all the way, allowing for the fact that one has not measured all of the visibilities, by letting some uncertainty into the model. The parameter which slows and stops the convergence is known as a regularisation parameter, and mathematically there are many different functions suitable for this role. In the maximum entropy method the entropy of the model map is used as the regularisation parameter. This can be seen in equation \ref{jeqn}

\begin{equation}
\label{jeqn}
J=H(I_{m})-\alpha\chi^{2}(V_{m},V_{d})
\end{equation}

\noindent
 where $J$ is the function maximised to deconvolve the map, $\chi^{2}$ is a measure of the difference between the model visibilities $V_{m}$ and the data visibilities $V_{d}$ and $H(I_{m})$ is the entropy of the model map. $\alpha$ is a Lagrangian parameter that balances the importance of maximising entropy, which represents uncertainty due to noise and the missing visibilities, against minimising $\chi^{2}$, which represents the fidelity of the model to the data that has actually been observed. The maximisation of $J$ is carried out in small steps allowing the real data features to appear if the dirty map strongly demands them. Additional terms can also be added to equation \ref{jeqn} to represent additional constraints, such as a constraint that the total flux of the model be equal to the observed flux.\\

\noindent
MEM also has the interesting property of ``super-resolution'', meaning that the algorithm can continue to converge to a factor of four better than the maximum resolution possible according to the sampling theorem of Fourier theory (Eq. \ref{reseqn}) given by:

\begin{equation}
\label{reseqn}
x_{min}=\frac{1}{u_{max}} \hspace{3cm} y_{min}=\frac{1}{v_{max}}
\end{equation}

\noindent
Eq. \ref{reseqn} represents the smallest scale at which the map can contain real data, thus the convergence of MEM to a resolution lower than this is a mathematical effect and data at such high resolution may not represent real features. However as MEM models sources as continuous distributions, rather than as the set of delta functions used in the standard CLEAN algorithm, it is possible to convolve the MEM model with a beam appreciably smaller than the corresponding CLEAN beam and yield reliable features at resolutions higher than is possible with standard CLEAN. This is potentially of considerable interest when analysing structure in regions that are only marginally resolved with CLEAN.  Discussions of MEM and implementations of it can be found in \cite{cornwellevans}, \cite{holdaway}, \cite{gullskilling}, \cite{sault}.

\section{Choice of the form of entropy}

A common form of entropy used in image analysis is the Shannon entropy:

\begin{equation}
\label{shannoneqn}
H=-\sum_{k} I_{k}(log(\frac{I_{k}}{IB_{k}}))
\end{equation}

\noindent
where $H$ is the entropy of the entire map, $I_{k}$ represents the $k^{th}$ pixel of the Stokes I model map and $IB_{k}$ represents the $k^{th}$ pixel of a bias map (normally taken to be a flat map with total flux equal to the total observed flux). This form of entropy is implemented widely and is available in many popular software packages, including in AIPS as the task ``VTESS'' and in the CASA toolkit. However this form of entropy is unsuitable for studies of polarised emission as the Stokes Q and U maps may take on negative values. To rectify this, a form of entropy suitable for the deconvolution of polarisation data was developed by Gull and Skilling \cite{gullskilling}.

\begin{equation}
H=-\sum_{k} I_{k}(log(\frac{I_{k}}{IB_{k}e})+\frac{1+m_{k}}{2}log(\frac{1+m_{k}}{2})+\frac{1-m_{k}}{2}log(\frac{1-m_{k}}{2}))
\end{equation}

\noindent
This form of entropy is very similar to Shannon entropy, except for the addition of some new terms involving $m_{k}$, the fractional polarisation of the model map at pixel $k$. As Gull and Skilling entropy is not widely implemented in current software packages, a C++ program was written to perform the full polarisation deconvolution. This program was based largely on MIRIAD's ``PMOSMEM'' task (author: R. Sault) \cite{miriad} and AIPS's ``VTESS'' task (author: T. J. Cornwell) \cite{aips}. If simultaneous MEM polarisation images are available at several frequencies they can be used to construct a RM map in the usual way. The first results of this study are presented below.
\vspace{-5mm}
\section{Results}

\begin{figure}

  \subfigure[CLEAN Rotation Measure map]{
    \includegraphics[scale=0.3]{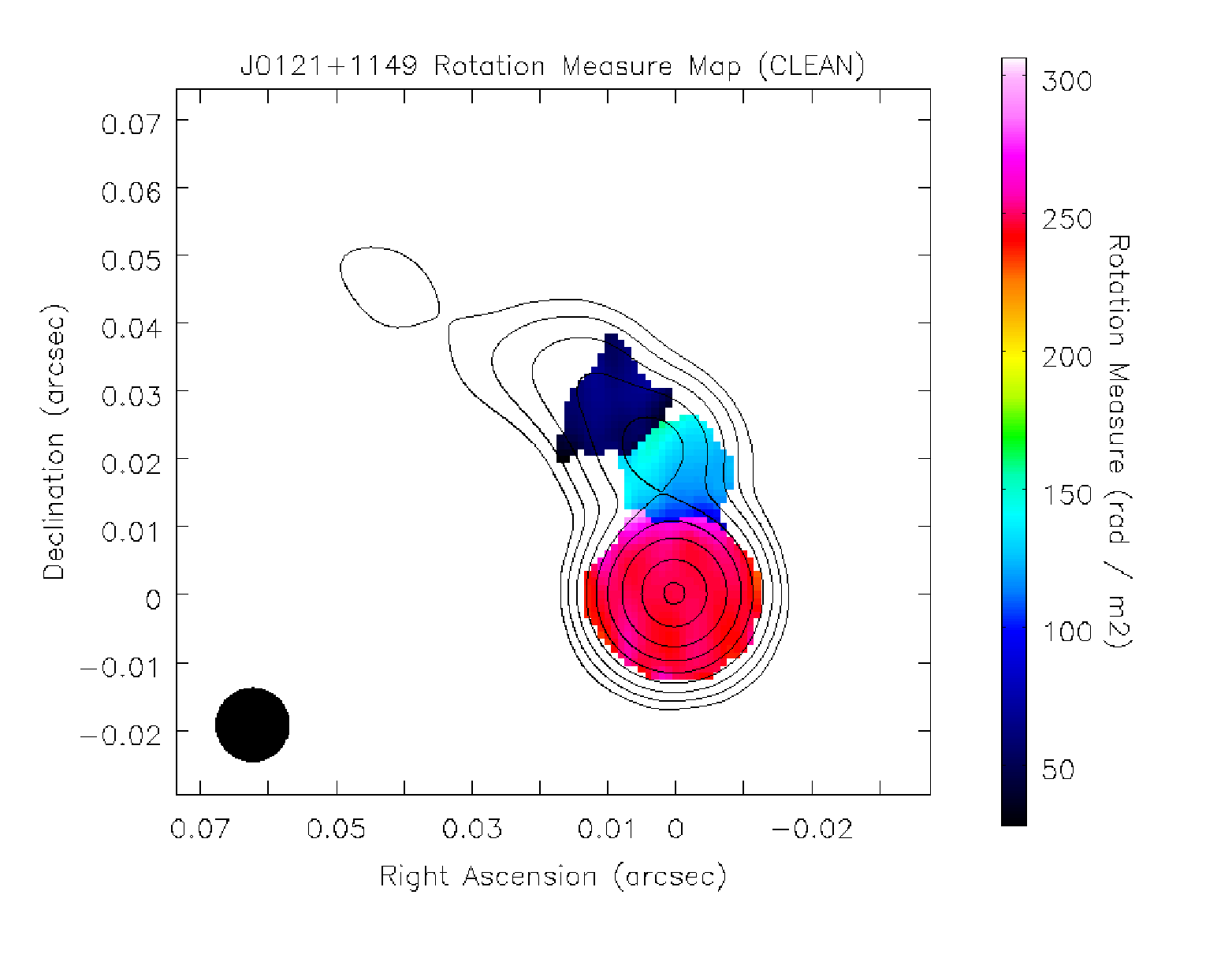}
  }
  \hfill
  \subfigure[CLEAN intrinsic polarisation map]{
    \includegraphics[scale=0.3]{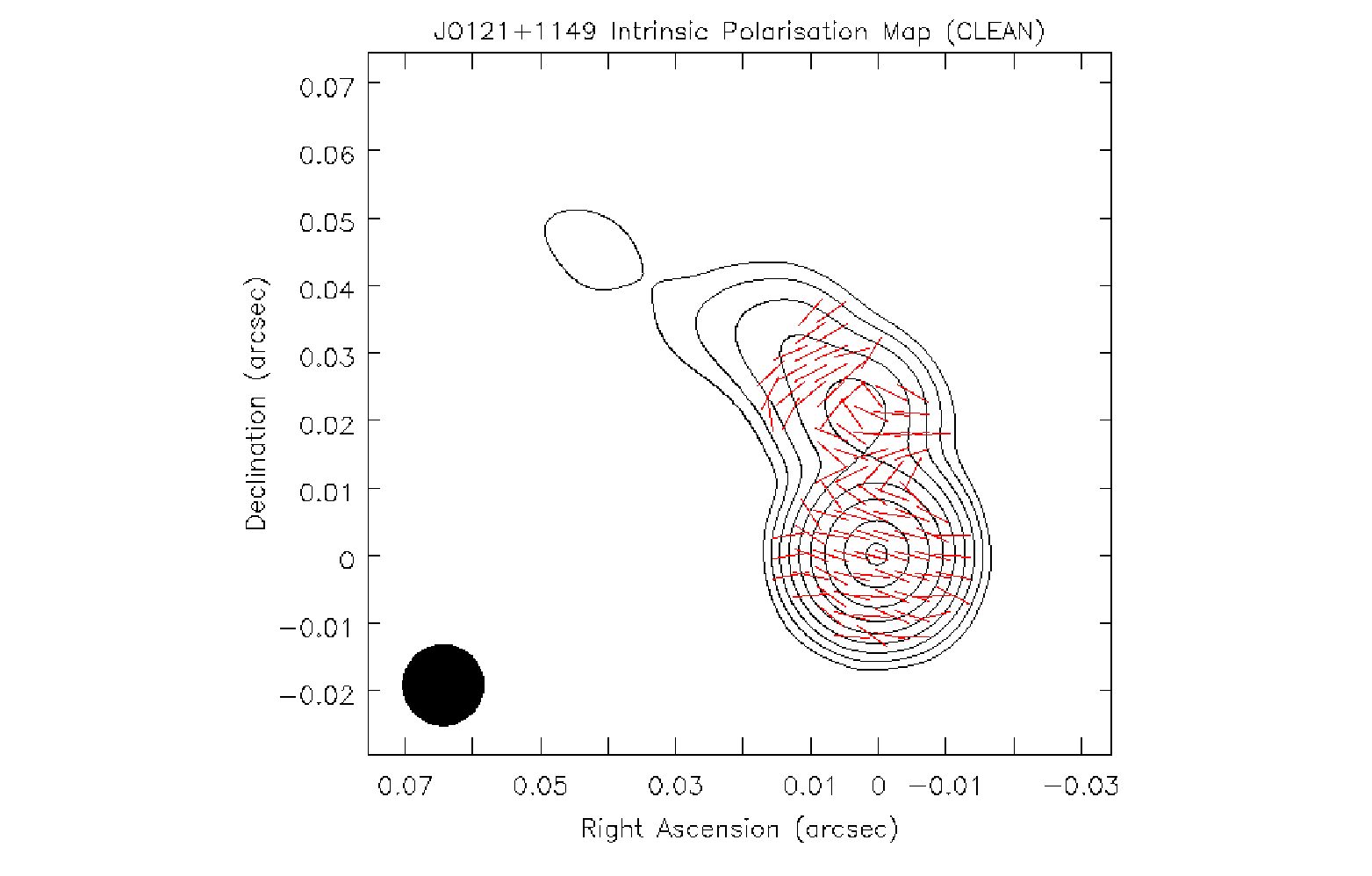}
  }
  \hfill
  \subfigure[MEM Rotation Measure map (CLEAN beam)]{
    \includegraphics[scale=0.3]{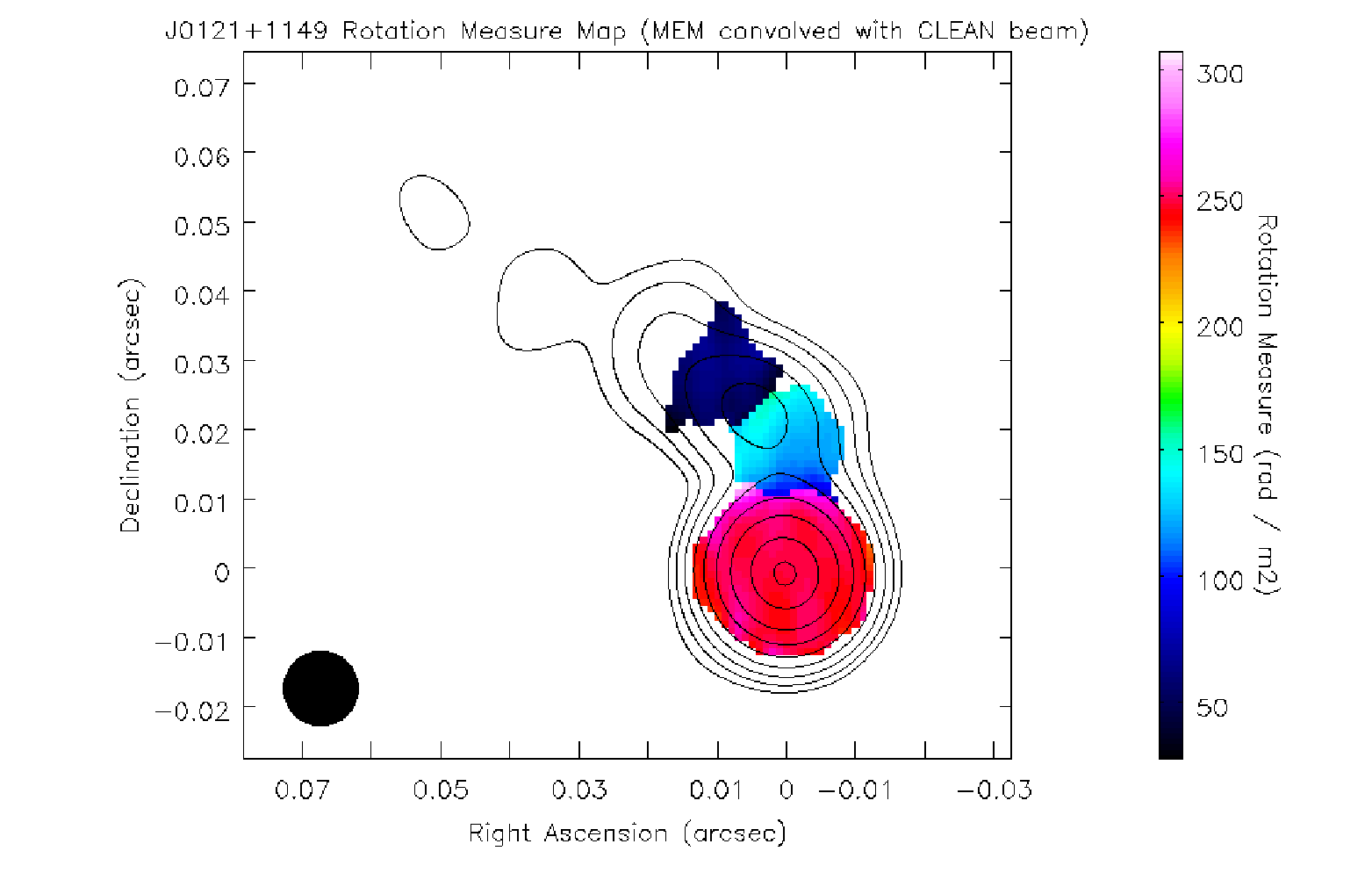}
  }
  \hfill
  \subfigure[MEM intrinsic polarisation map (CLEAN beam)]{
    \includegraphics[scale=0.3]{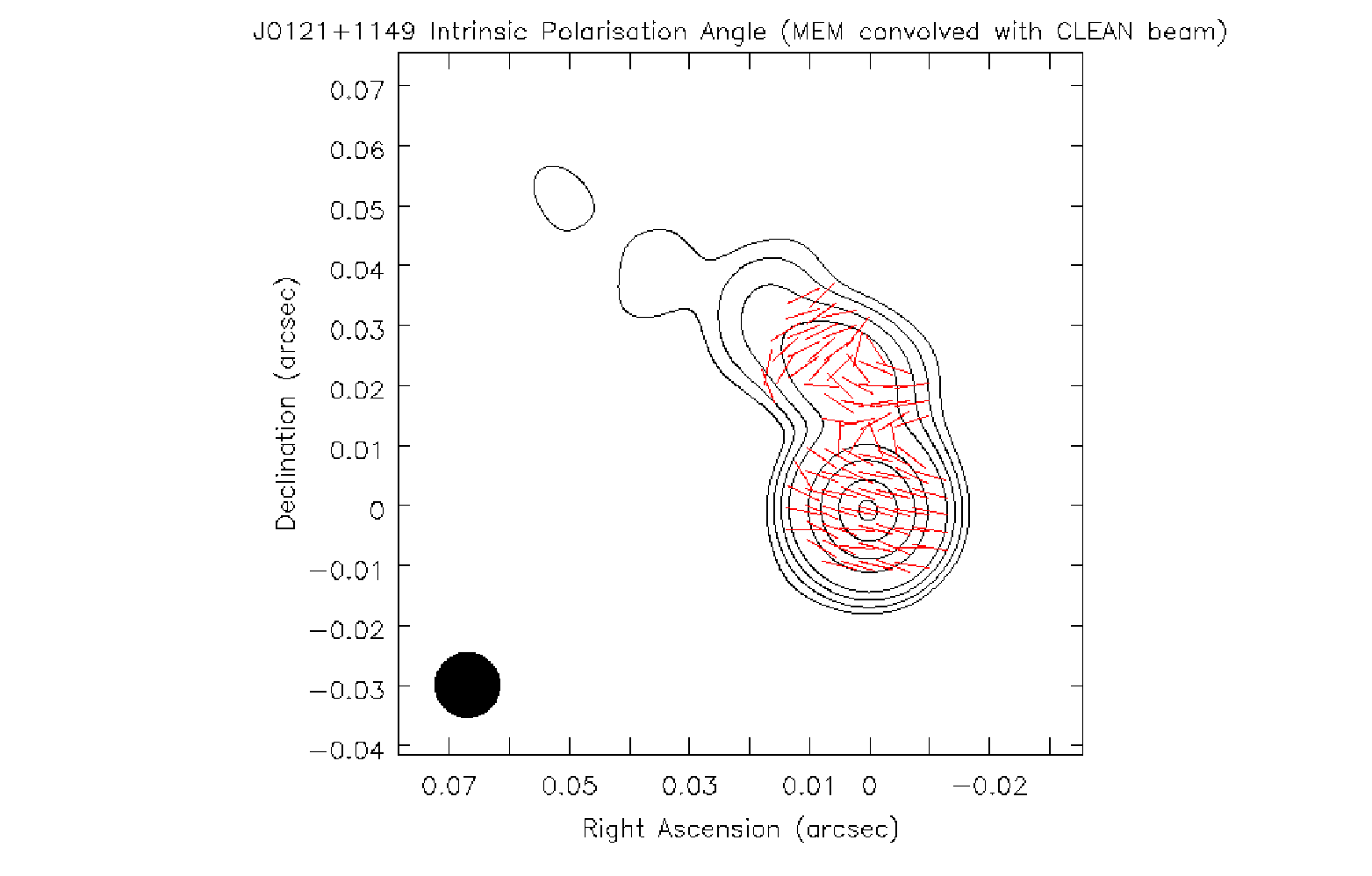}
  }
  \hfill
  \subfigure[MEM Rotation Measure map (1/3 CLEAN beam)]{
    \includegraphics[scale=0.3]{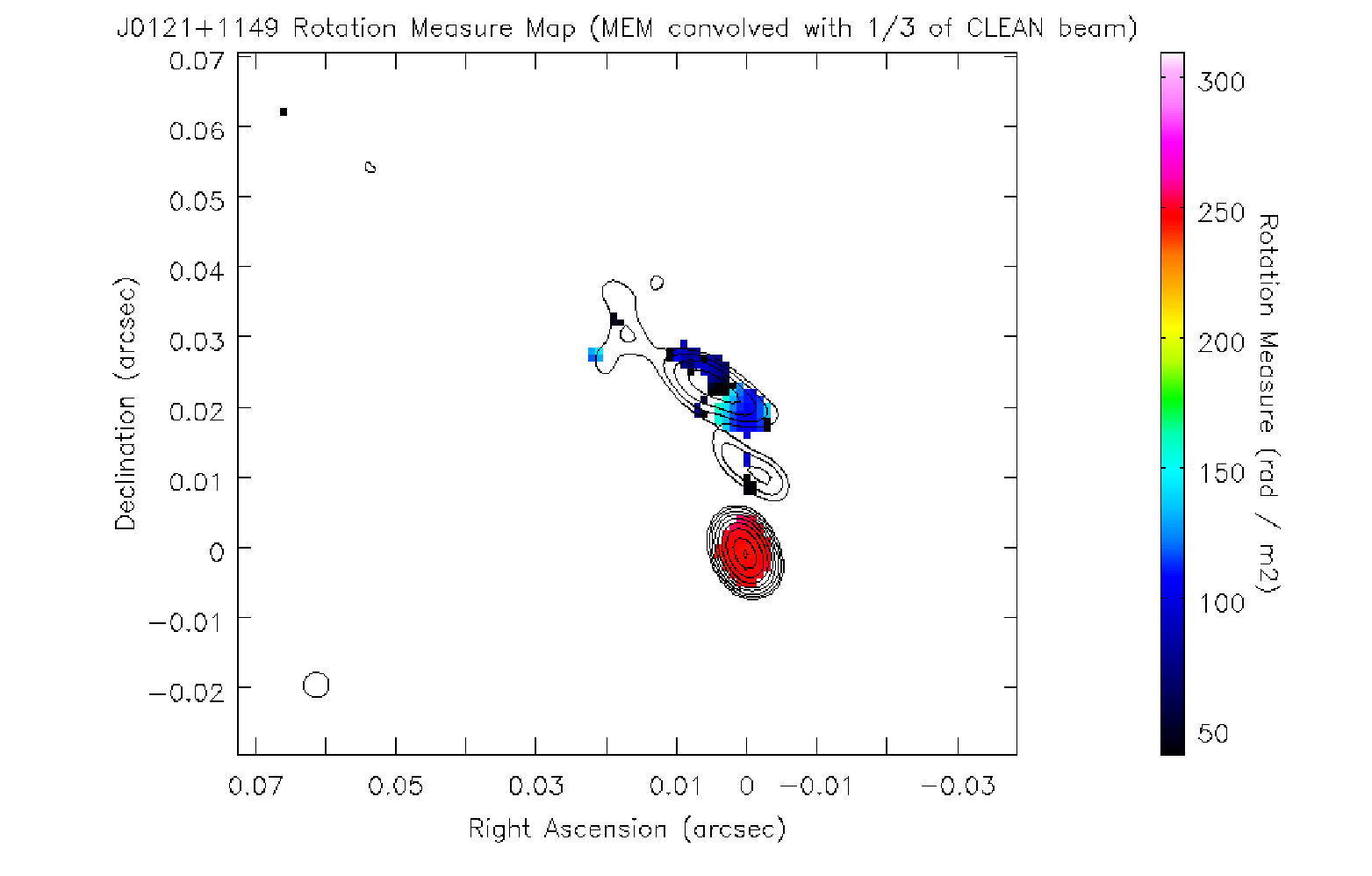}
    \label{J0121_MEM_TB}
  }
  \hfill
  \subfigure[MEM intrinsic polarisation map (1/3 CLEAN beam)]{
    \includegraphics[scale=0.3]{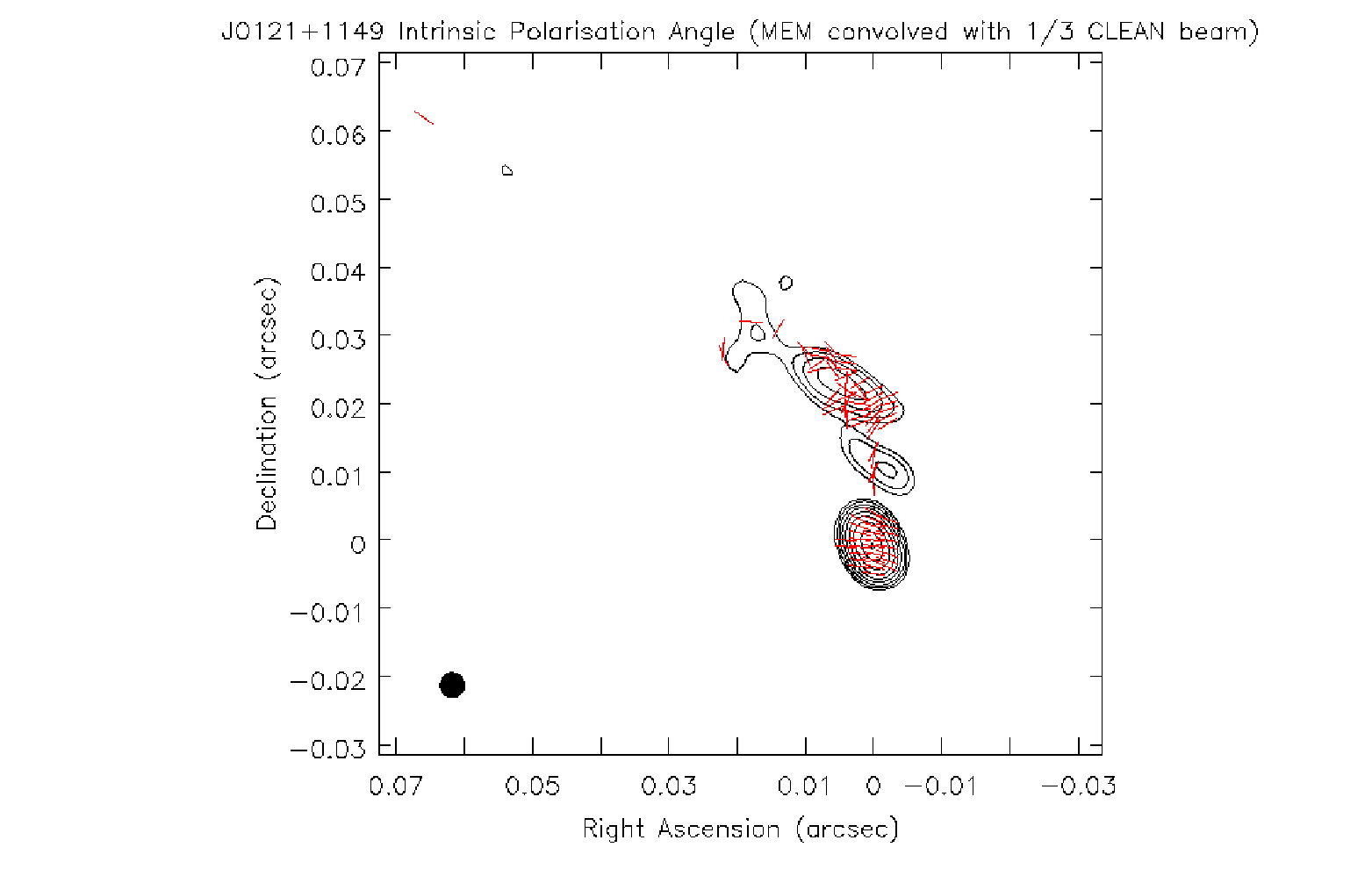}
    \label{J0121_MEM_TB_pol}
  }
  
  \caption{Rotation Measure and intrinsic polarisation maps of J0121+1149. The contours represent Stokes I emission at 1358 MHz. Contours shown are 0.5, 1 , 2, 4, 8, 16, 32, 64, 95 \% of peak. Maps (a) and (b) have a peak of 0.82 Jy/Beam. Maps (c) and (d) have a peak of 0.80 Jy/Beam. Maps (e) and (f) have a peak of 0.38 Jy/Beam.}
  \label{j0121_plots}

\end{figure}
\noindent
{\bf J0121+1149:} Multi-wavelength observations of the source at 1358, 1430, 1493 and 1665 MHz were made in 2010 \cite{coughlan}. Figure \ref{j0121_plots} shows CLEAN and MEM rotation measure and intrinsic polarisation maps for J0121+1149 made from these data. Each map shows 3 distinct regions of Faraday rotation. The CLEAN and MEM maps convolved with the full CLEAN beam show excellent agreement in the location and magnitude of these features. The use of a second algorithm to confirm these features increases confidence that they represent real structure. One can also convolve the MEM map with a beam somewhat smaller than the CLEAN beam to examine the source at a higher resolution. Figs. \ref{J0121_MEM_TB} and \ref{J0121_MEM_TB_pol} show MEM maps convolved with a beam with a size 1/3 of the usual CLEAN beam. An additional jet component in the intensity emission is visible at this resolution, however there is no obvious sign of any relationship between the sudden change in RM value and intensity components in the jet.\\

\begin{figure}
  
  \subfigure[CLEAN Rotation Measure map]{
    \includegraphics[scale=0.3]{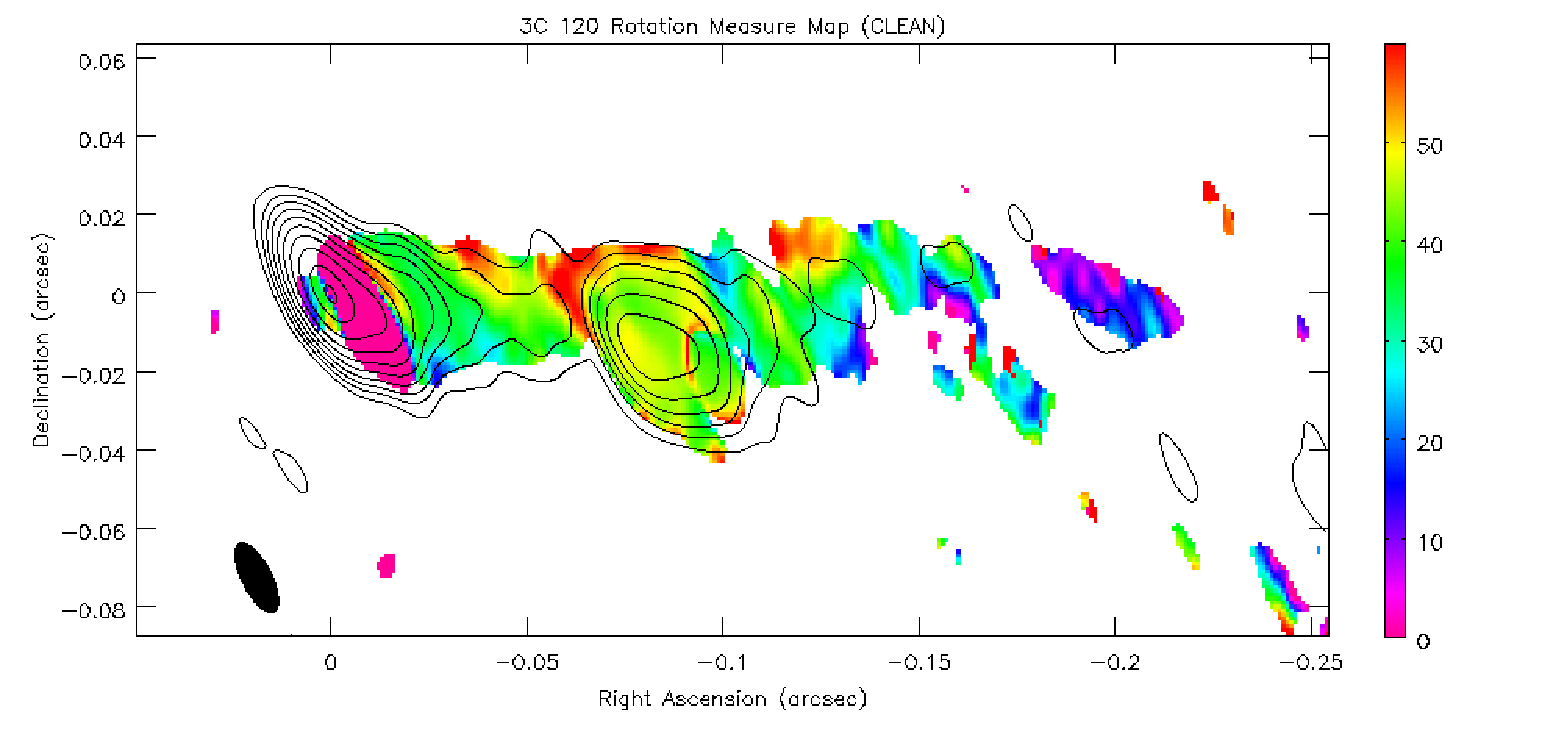}
  }
  \hfill
  \subfigure[CLEAN intrinsic polarisation map]{
    \includegraphics[scale=0.3]{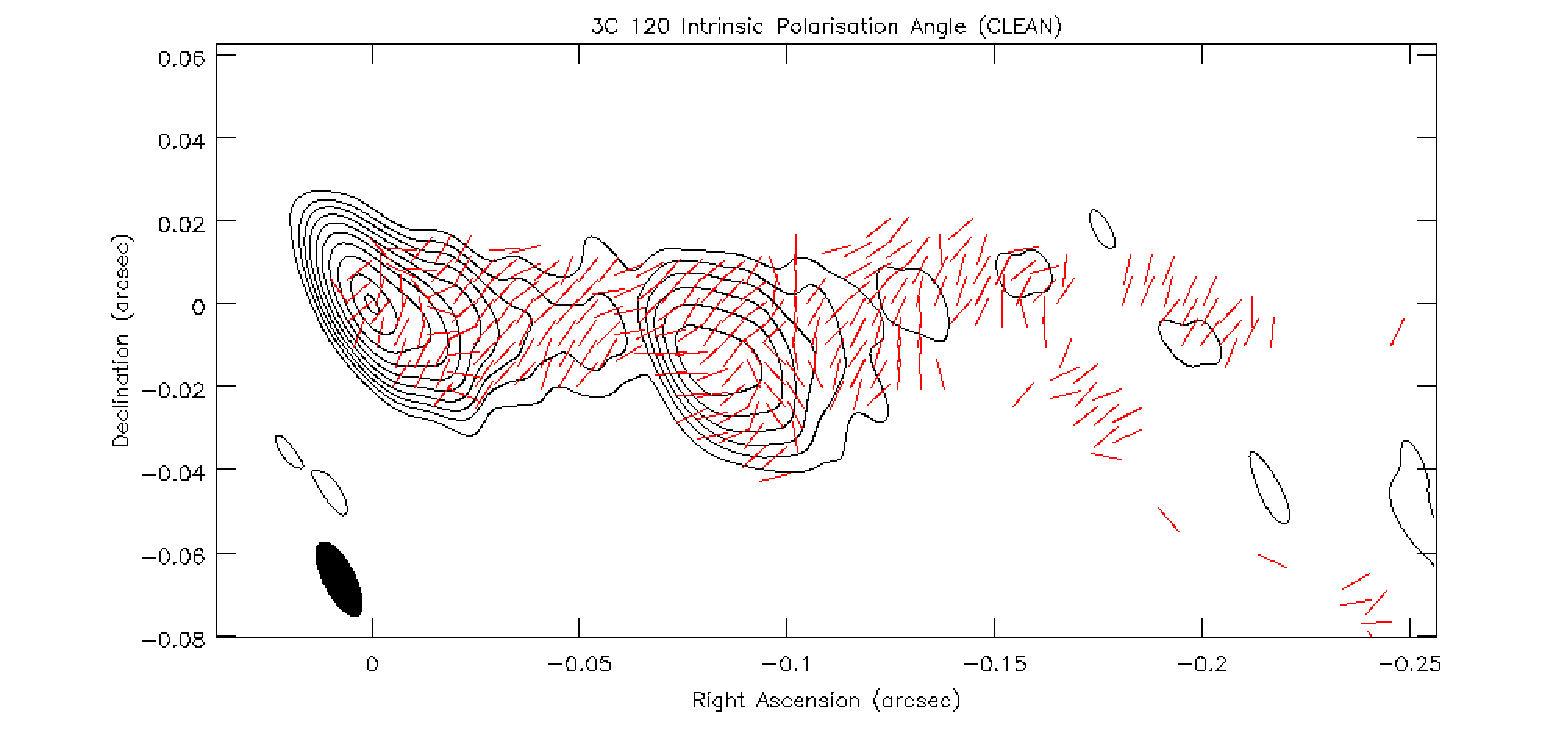}
  }
  \hfill
  \subfigure[MEM Rotation Measure map (1/2 CLEAN beam)]{
    \includegraphics[scale=0.3]{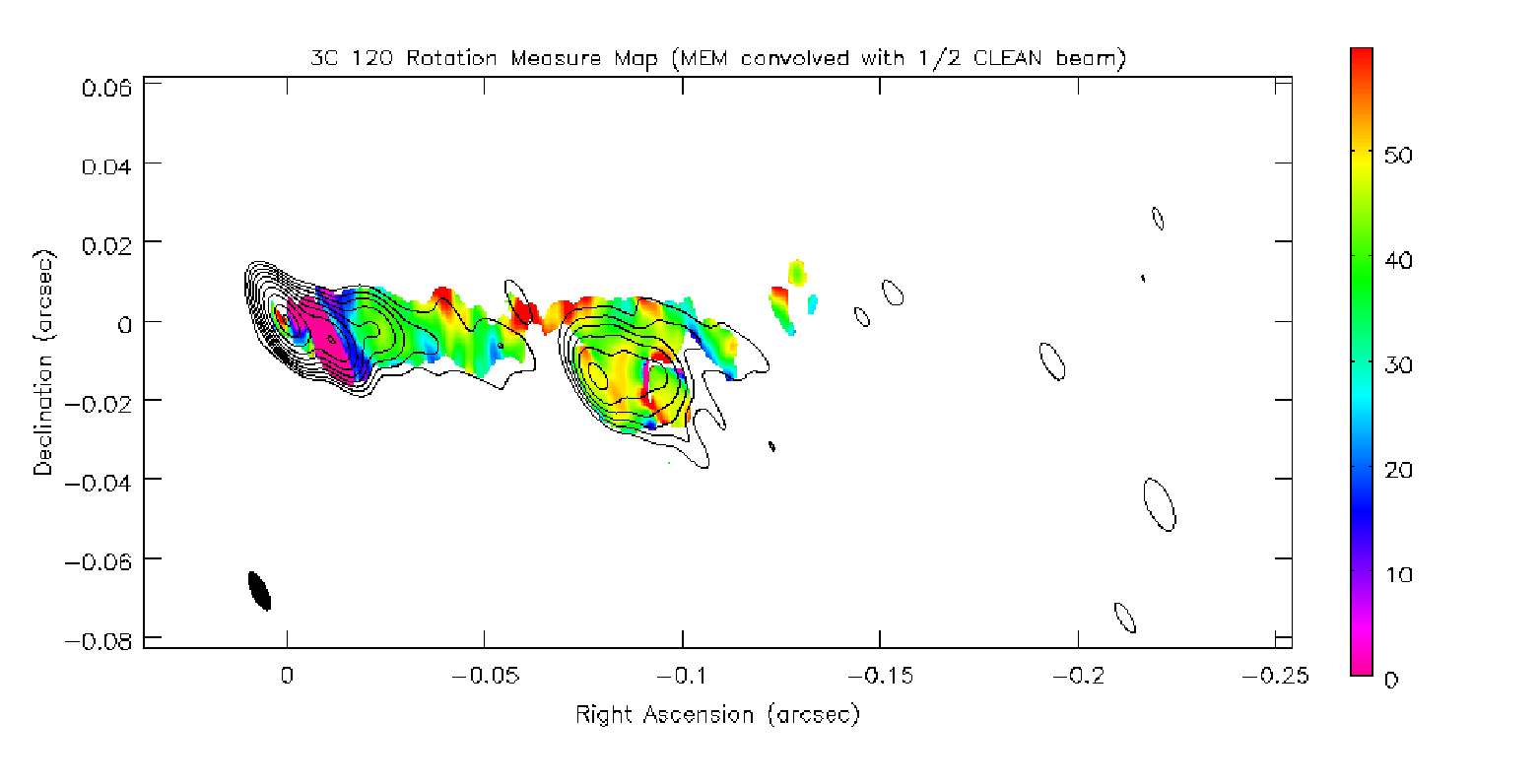}
  }
  \hfill
  \subfigure[MEM intrinsic polarisation map (1/2 CLEAN beam)]{
    \includegraphics[scale=0.3]{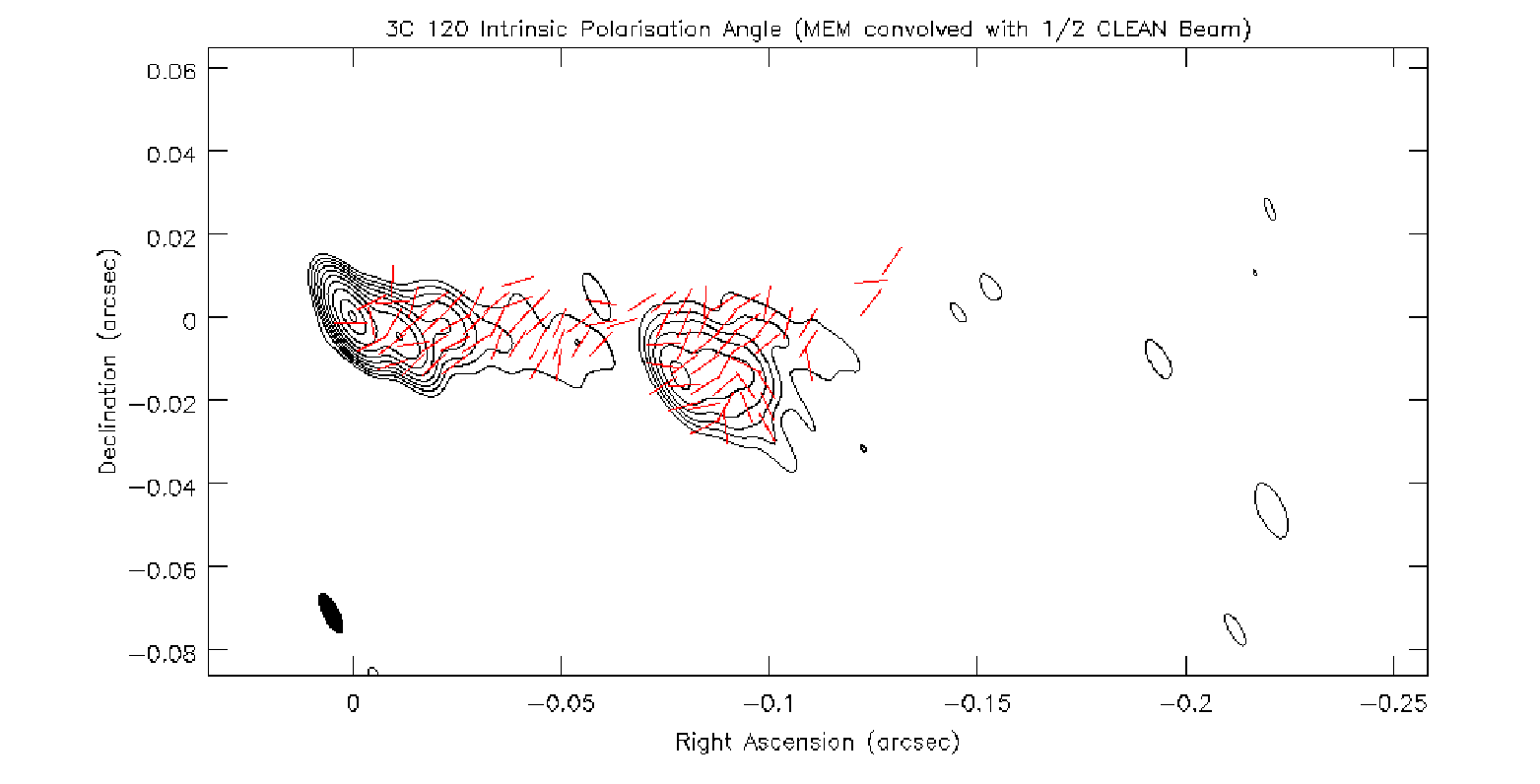}
  }
  
  \caption{Rotation Measure and intrinsic polarisation maps of 3C 120. The contours represent Stokes I emission at 1358 MHz.  Contours shown are 0.5, 1 , 2, 4, 8, 16, 32, 64, 95 \% of peak. Maps (a) and (b) have a peak of 1.36 Jy/Beam. Maps (c) and (d) have a peak of 0.95 Jy/Beam.}
   \label{3c120_plots}

\end{figure}

\noindent
{\bf 3C 120:} Observations of 3C 120 were also made as part of the same experiment  \cite{coughlan}. 3C 120 possesses an extended jet with a rich Faraday rotation structure (\cite{gomez}, Figs. 5 \& 6), visible in Figure \ref{3c120_plots}. There is good agreement between features seen in the CLEAN and MEM Faraday rotation and intrinsic polarisation maps, even though the MEM maps have been made had been made with twice the resolution of the CLEAN maps. This high resolution MEM map shows the local jet direction more clearly, and can be used with the CLEAN map to search for signs of interesting Faraday rotation behaviour, such as transverse gradients which may indicate the presence of a helical magnetic field. In the case of 3C 120 there are some regions of both the CLEAN and high resolution MEM maps that show hints of such gradients, but also large regions where the RM structure appears to be random, suggesting the possible presence of inhomogeneous Faraday screens between between the source and the observing telescopes.

\vspace{-5mm}
\section{Conclusions}

We have written a new C++ code to carry out a full MEM polarisation deconvolution of VLBI data. The MEM code has been successfully applied to multi-frequency VLBI polarisation data and shows good agreement with CLEAN images when convolved with a similar beam. MEM's enhanced resolution over CLEAN can potentially provide a more complete understanding of fine structure in VLBI polarisation images of AGN. Further work will involve using Monte Carlo simulations to verify the accuracy of the MEM algorithm as implemented in C++ and to characterise any systematic and random errors that may be present.

\vspace{-5mm}
\section*{Acknowledgements}

\noindent
This work was supported by the Irish Research Council for Science, Engineering and Technology (IRCSET).

\end{document}